\documentstyle[epsf,12pt]{article}
\begin{document}
\setcounter{page}{0}
\thispagestyle{empty}
\begin{flushright}
GUTPA/95/01/1
\end{flushright}
\vskip .2in
\begin{center}
{\large\bf Could There Be a Fourth Generation of
Quarks Without More Leptons?\\}
\vskip .4cm
{\bf C. D. Froggatt$^1$, D. J. Smith$^1$ and H. B. Nielsen$^2$}
\vskip .2cm
{ $^1$ \em Department of Physics and Astronomy,
University of Glasgow,\\ Glasgow G12 8QQ, UK}
\vskip .2cm
%{\bf $\rm{ and H. B. Nielsen^2}$}
%\vskip .2cm
{$^2$ \em Niels Bohr Institute, Blegdamsvej 17-21,
\\DK 2100 Copenhagen, Denmark}
\end{center}
\section*{ }
\begin{center}
\vskip  .4cm
{\large\bf Abstract}
\end{center}
We investigate the possibility of adding a
fourth generation of quarks. We also
extend the Standard Model gauge group
by adding another $SU(N)$ component. In
order to cancel the contributions of
the fourth generation of quarks to the
gauge anomalies we must add a generation
of fermions coupling to the $SU(N)$
group. This model has many features
similar to the Standard Model and, for
example, includes a natural generalisation
of the Standard Model charge
quantisation rule. We discuss the
phenomenology of this model and, in
particular, show that the infrared
quasi-fixed point values of the Yukawa
coupling constants put upper limits
on the new quark masses close to the
present experimental lower bounds.
\vskip 4.5cm

\section{Introduction}
\label{Intro}

The LEP determinations of the invisible
partial decay width of the $Z^0$
gauge boson show there are just three
neutrinos of the usual type
with mass less than $M_{Z}/2$.
This result is naturally interpreted
to imply that there are just three
generations of quarks and leptons.
In this letter we consider the
possibility that there is an extra
generation of quarks without accompanying leptons. There is of
course a well-known problem with such a "half generation"
consisting only of quarks: the gauge anomalies
associated with triangle
diagrams, having electroweak gauge bosons (but no QCD gluons)
attached to the vertices, would not cancel.
So some further particles would have to be added to the model
in order to cancel these anomalies. At first one could worry
that it might be difficult to cancel all the anomalies
in a simple way. However we shall argue that the
uncancelled quark anomalies make up
a natural unit that is rather easily compensated by a kind of
techniquark, or rather technilepton since it does not couple to
$SU(3)_{C}$, generation. Our philosophy
in the present article is to seek extensions
of the Standard Model (SM)
incorporating a fourth quark generation
and retaining as many of the
distinctive properties of the SM gauge
group as possible. We should
even like to argue that the proposed
fourth quark generation model
is the unique extension of the SM,
which retains most of the features
of the SM and is still (just) viable phenomenologically.

So we now summarise the characteristic properties of the Standard
Model Group (SMG). SMG is defined to be the group
$U(1) \otimes SU(2) \otimes SU(3)$ supplemented by the charge
quantisation rule:
\begin{equation}
\label{SMchqu}
\frac{y}{2}+\frac{1}{2} d+\frac{1}{3} t
        \equiv 0\pmod{1}
\end{equation}
The duality $ d$ has value 1 if
the representation is an $SU(2)$ doublet and 0
if it is an $SU(2)$ singlet.
The triality $t$ has value 1 if the
representation is an $SU(3)$ triplet, 0 if
it is an $SU(3)$ singlet, and -1 if it
is an $SU(3)$ anti-triplet. The
imposition of the charge quantisation
rule eq.~(\ref{SMchqu}) corresponds
to dividing out an appropriate discrete group D:
\begin{equation}
\label{SMGroup}
SMG \equiv U(1) \otimes SU(2) \otimes SU(3) /D
\end{equation}
In fact SMG is the proper subgroup \cite{OrigOfSym,ORaif}
$S(U(2) \otimes U(3))$ of $SU(5)$.
The SM charge quantisation rule eq.~(\ref{SMchqu})
implies that the
value of the weak hypercharge $y/2$ of
an irreducible representation of
SMG determines uniquely the duality and
triality of the representation.

We wish to consider extensions of SMG with
the same distinctive features
as SMG itself and are thereby led
to add another $SU(N)$ factor to
the sequence $U(1) \otimes SU(2) \otimes SU(3)$,
giving gauge groups of the type:
\begin{equation}
\label{SMG23NGroup}
SMG_{2,3,N} \equiv U(1) \otimes
SU(2) \otimes SU(3) \otimes SU(N) / D_{N}
\end{equation}
Here $D_{N}$ is a discrete group which ensures that a generalised
charge quantisation rule of the form:
\begin{equation}
\label{GenChQuant}
\frac{y}{2}+\frac{1}{2}d+\frac{1}{3}t
+\frac{m_{\scriptscriptstyle{N}}}{N}n \equiv 0 \pmod 1
\end{equation}
is satisfied, where $n$ is the N-ality
of the representation and $m_N$
is an integer which is not a multiple of N.

Another feature that we take over from the SM is the
principle of using only small (fundamental or singlet) fermion
representations; a feature not shared with
supersymmetric models, where the
gauginos are in adjoint representations.
We note that N-ality has value 1
if a representation is an $SU(N)$
N-plet, 0 if it is an $SU(N)$ singlet, and -1
if it is an $SU(N)$
anti-N-plet. In order that the generalised
charge quantisation rule,
eq. (\ref{GenChQuant}), continues to determine
the duality, triality
and N-ality of an irreducible representation
once $y/2$ is specified,
2, 3 and N must be mutually prime.

We will consider the fundamental scale to be
the Planck mass ($M_{Planck}$)
and our models will be a full description
of physics without gravity
below this scale. By requiring an anomaly
free theory and the absence of
any Landau poles below $M_{Planck}$,
we obtain an essentially unique
extension of the SM with an extra
generation of quarks, but with the
extra generation of leptons replaced
by fermions in the fundamental
representations of the $SU(N)$ component of $SMG_{2,3,N}$.

We assume that the extra $SU(N)$ component of
the gauge group is not
spontaneously broken and therefore confines
forming fermion condensates.
As we already know
from the SM, the $SU(3)$ component acts as
a technicolour group \cite{TC} and
gives a contribution to the $W^{\pm}$
and $Z^{0}$ masses. In the SM this
contribution is very small but when
confining groups with $N>3$ are considered
we must carefully
consider the effect this will have.
Since we are not wanting the complications
of extended technicolour in order to generate
quark and lepton masses,
we assume that there is a Higgs doublet and
that the masses of the weak
gauge bosons are generated by a combination
of the Higgs sector of the theory
and the technicolour effects of the gauge groups.
Thus the extra $SU(N)$
component of the gauge group acts only as a partial
technicolour \cite{TechniHiggs}. It follows that the $SU(N)$
confinement scale must be at the TeV scale or below.

In order to avoid conflict with precision
electroweak data, it is not
phenomenologically consistent to introduce
a large number of extra
$SU(2)_{L}$ doublets. We shall therefore concentrate on the
phenomenology of the $SMG_{2,3,5}$ model which has
the minimal value of $N=5$.

\section{Structure of the $SMG_{2,3,N}$ Model}
The three SM generations form representations
of $SMG_{2,3,N}$ and we now
wish to consider a new ``generation'' of fermions coupling to the
$SU(N)$, but not the $SU(3)$, component of the group.
We assume all the fermions get a mass
via the SM Higgs mechanism and, by analogy with the SM quarks, we
consider a generation of $SU(N)$ ``quarks''
made up of the left-handed
fermion representation $(y,2,N)$ together
with the left-handed anti-fermion
representations $(-[y+1],1,\overline{N})$
and $(-[y-1],1,\overline{N})$.
The weak hypercharge $y$ must of course satisfy the charge
quantisation rule eq.(\ref{GenChQuant}).
We now consider the cancellation
of the anomalies generated by such
a generation of $SU(N)$ ``quarks''.

The contributions to each type of anomaly
from this generation of $SU(N)$
``quarks'' are:
\begin{displaymath}
\begin{array}{lllll}
[SU(N)]^{3} & \rightarrow &  &  & 0 \\
\left[SU(N)\right]^{2}U(1) & \rightarrow & 2y-(y+1)-(y-1)
& = & 0 \\
\left[Grav\right]^{2}U(1) & \rightarrow & 2y-(y+1)-(y-1)
& = & 0 \\
\left[U(1)\right]^{3} & \rightarrow
& N[2y^{3}-(y+1)^{3}-(y-1)^{3}]
& = & -6Ny \\
\left[SU(2)\right]^{2}U(1) & \rightarrow & & & Ny
\end{array}
\end{displaymath}
So we must cancel the resulting $\left[SU(2)\right]^{2}U(1)$ and
$\left[U(1)\right]^{3}$ anomalies against
other fermion representations
in order to obtain a consistent theory.
We can in fact cancel the anomalies
against a fourth generation of $SU(3)_{C}$ quarks,
in just the same way
that the quark anomalies cancel the lepton
anomalies in the three SM
generations \footnote{The resulting
generation of $SU(3)$ quarks and
$SU(N)$ ``quarks'' is the smallest mass-protected anomaly free
representation which couples non-trivially to all the components
of the gauge group $SMG_{2,3,N}$. In the technicolour literature
these $SU(N)$ ``quarks'' are usually
called technileptons as they do
not couple to $SU(3)_{C}$.}.

The non-zero anomalies due to a fourth generation of quarks are:
\begin{displaymath}
\begin{array}{lllll}
\left[U(1)\right]^{3} & \rightarrow & 3[2(\frac{1}{3})^{3} -
(\frac{4}{3})^{3} + (\frac{2}{3})^{3}] & = & -6 \\
\left[SU(2)\right]^{2}U(1) & \rightarrow &
3 \cdot \frac{1}{3} & = & 1
\end{array}
\end{displaymath}
Anomaly cancellation between a generation
of $SU(N)$ ``quarks'' and
a fourth generation of $SU(3)_C$ quarks gives the condition:
\begin{equation}
\label{anomcancel}
Ny+1=0
\end{equation}
and hence
\begin{equation}
y=-\frac{1}{N}
\end{equation}
However the weak hypercharge $y$ of the
$SU(N)$ ``quarks'' must satisfy
the charge quantisation rule eq.(\ref{GenChQuant}),
which takes the form:
\begin{equation}
\label{SUNhyperch}
y = 2J - 1 - \frac{2m_{\scriptscriptstyle{N}}}{N}
\end{equation}
where $J$ is an integer. So we have a solution with $J = 1$ and
\begin{equation}
\label{mNvalue}
m_N = \frac{1}{2}(N+1)
\end{equation}
Table~\ref{SU5Quarks} shows the properties of the left-handed
fermions belonging to such a generation of
$SU(N)$ ``quarks'' for the
phenomenologically interesting case: N=5.
Finally we remark that $N$ is odd and hence
Table~\ref{SU5Quarks} contains
an odd number of $SU(2)$ doublets.
Combined with the three $SU(2)$ doublets
in a fourth generation of quarks,
this gives an even number of doublets and
hence there is no Witten discrete $SU(2)$
anomaly \cite{WittenAnom}.

\section{Phenomenology of the $SMG_{2,3,5}$ Model}
We shall now discuss the minimal extension of
the SM in our class of models.
It is based on the gauge group
$SMG_{2,3,5} \equiv U(1) \otimes
SU(2) \otimes SU(3) \otimes SU(5) / D_{5}$
and, in addition to the three
SM generations, contains a fourth generation of quarks and a
single generation of $SU(5)$ ``quarks'' as specified in
Table~\ref{SU5Quarks}.

We assume that the $SU(5)$ component of
the gauge group confines and
that the $SU(5)$ ``quarks'' form condensates.
These condensates have the
same quantum numbers as the SM Higgs boson
and contribute to the
$W^{\pm}$ and $Z^0$ masses, via the usual
technicolour \cite{TC} mechanism.

We stress that we are not proposing a
technicolour model as such, but simply
taking into account the unavoidable
effect that adding an $SU(5)$ group has.
We are assuming that the Higgs sector
of our models is the same as in the SM,
i.e.\ one Higgs doublet. Then the VEV
due to the Higgs field, $<\phi_{WS}>$, is
related to the total VEV, $v$, and the
contribution from $SU(5)$ due to
fermion condensates, $F_{\pi_N}$,
by the relation
\begin{equation}
<\phi_{WS}>^2 + F_{\pi_N}^2 = v^2 = (246 \; \rm{GeV})^2
\end{equation}
which is exactly the same as in
technicolour models with a scalar \cite{TechniHiggs}.

The fermion running masses, $m_f$, are related
to the Higgs field VEV in the usual way;
\begin{equation}
m_f = \frac{y_f}{\sqrt{2}}<\phi_{WS}>
\label{runmass}
\end{equation}
where $y_f$ is the Yukawa coupling constant for the fermion $f$
($y$ is used for
both Yukawa coupling and weak hypercharge
but it should be obvious from the
context which is being referred to).
In order to avoid any significant
suppression of the top quark and
other fermion masses, due to the reduction of
$<\phi_{WS}>$ below its SM value, we usually imagine taking
\begin{equation}
F_{\pi_N} \approx 75 \; \rm{GeV}
\end{equation}
and thus
\begin{equation}
<\phi_{WS}> \quad \approx \; 234 \; \rm{GeV}
\end{equation}
This leads to a 5\% reduction of the fermion
masses relative to the value if
there was no technicolour contribution. It also implies a typical
$SU(5)$ ``hadron'' mass \cite{georgi}, due to confinement, of
$\Lambda \simeq 400$ GeV.

We now briefly discuss the experimental
constraints on the masses of the new
fermions in the model. First we will consider the usual $SU(3)$
quarks and then the $SU(5)$ ``quarks''.

The top quark has recently been observed by the CDF and D0
\cite{CDFtop} collaborations, with a mass of $180 \pm 12$ GeV.
We use the dilepton mode analyses of the CDF and
D0 \cite{CDF4gen} groups to place a lower limit
on the possible masses of a
fourth generation of $t'$ and $b'$ quarks.
If the $t'$ quark is lighter
than the $b'$ quark, it is expected to decay via the mode
$t' \rightarrow bW^{+}$ and hence give a
dilepton signal similar to the
top quark. If the $b'$ quark is lighter than
the $t'$ and top quarks, it is
expected to decay via the mode $b' \rightarrow cW^{-}$
and again give a
similar dilepton signal. So we take the limit
on the pole masses of possible
fourth generation quarks, $t'$ and $b'$, to be
\begin{equation}
M_{t'}, M_{b'} > 130 \; \rm{GeV}
\end{equation}
from the dilepton analyses of the CDF and D0
groups \footnote{Different limits
apply if other decay modes are dominant \cite{gunion}}.

The above experimental limits do not apply to
possible $SU(5)$ ``quarks''.
These fermions would be more difficult to
produce and detect at hadron
colliders. As explained above, they would
anyway be expected to be confined
inside $SU(5)$ ``hadrons'' having masses of order
400 GeV. So we conclude that the $SU(5)$
``quarks'' would be unlikely
to be detected with current accelerators.

Precision electroweak data can be used to set
limits on possible new
physics at the electroweak scale.
For example, the closeness of the observed value of
the $\rho$ parameter
($\rho \equiv \frac{M_W^2}{M_Z^2 cos^2\theta_W}$)
to unity \cite{rho=1} indicates that the mass
squared differences within any new fermion
$SU(2)$ doublets must be small
($\ll (100 \; \rm{GeV})^2$).
We shall assume that these mass splittings
are small in our model.
More generally any new physics which affects
only the gauge boson self-energies can be
parameterised by the $S$, $T$ and $U$
parameters \cite{STU} or some equivalent
set of parameters \cite{Epsilons,Vs}.
The new physics contributions $T_{new}$ and $U_{new}$
to the $T$ and $U$ parameters in our model
can be neglected.

The $S$ parameter provides a strong constraint on the number of new
fermion doublets, since perturbatively a mass degenerate doublet
contributes $\frac{1}{6\pi}$ to $S$. Analysis of the precision
electroweak data gives \cite{langacker}:
\begin{equation}
\label{Snew}
S_{new} = -0.12 \pm 0.24
\end{equation}
for a Higgs mass $M_{H} \simeq 170$ GeV.
There are 8 new $SU(2)$ doublets in our $SMG_{2,3,5}$
model, made up
of a fourth generation of quarks and a generation
of $SU(5)$ ``quarks''. In
a single SM generation, there are
4 doublets. So treating all the new
fermions perturbatively,
their contribution to $S_{new}$ is equivalent to
that of two SM generations:
$S_{new} = \frac {4}{3\pi} \sim 0.42$. Thus,
perturbatively, our model deviates by $\sim$ 2 standard
deviations from the experimental value of $S_{new}$.
Non-perturbative corrections, estimated
by analogy with QCD, tend to increase the
$SU(5)$ fermion contribution to $S$
\cite{lane} by a factor $\sim$ 2 and raise the
deviation of our model
from experiment to $\sim$ 3 standard deviations.
Our model is therefore
just consistent with the precision electroweak data.

We now investigate the consistency of the model assuming it
to be valid up to the Planck scale.
In particular we require the absence of
Landau poles below $M_{Planck}$.

\begin{figure}
\epsfxsize=\textwidth
\epsffile[100 100 460 500]{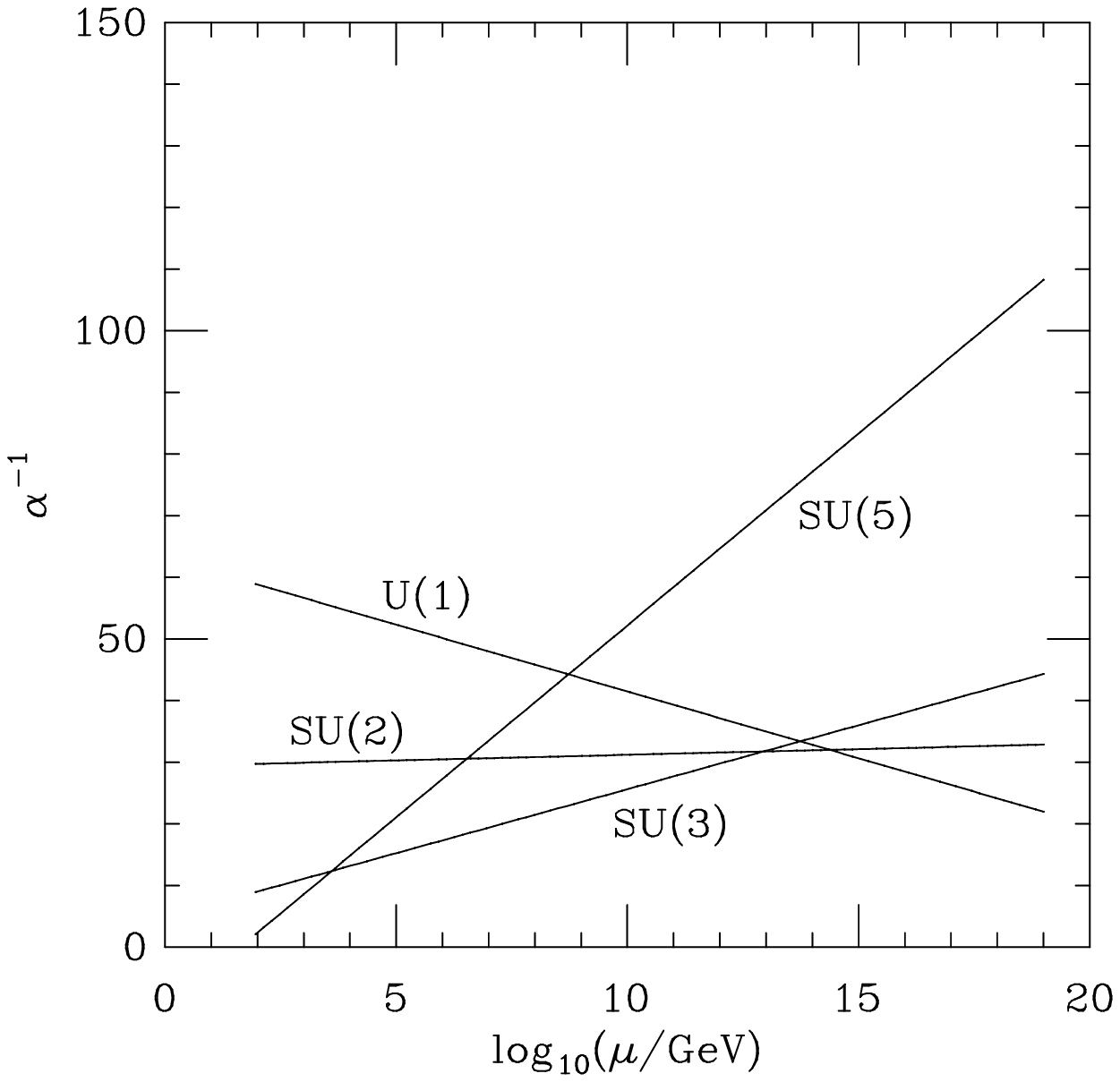}
\caption{}
%{$\alpha^{-1}$ from $M_Z$ to the Planck scale for each group.
%There are obviously no Landau poles so this model is self-consistent.}
\label{alph4g}
%\special{psfile=alph4g.ps angle=270 }
\end{figure}

Using the renormalisation group equations
(RGEs), we first examine how the gauge coupling constants vary
with energy in our model with 1 Higgs doublet, 3 SM generations,
a fourth quark generation and a generation of $SU(5)$ ``quarks''.
For convenience, we set the thresholds
for all the unknown fermions (4th generation
quarks and fermions coupling to
$SU(5)$), as well as for the top quark
and Higgs boson, to $M_Z$. The absence
of Landau poles in this case will
guarantee their absence if some of the
thresholds are set higher than $M_Z$.
There are four fine structure constants, which we shall label by
$\alpha_{1}$, $\alpha_{2}$, $\alpha_{3}$ and
$\alpha_{5}$, corresponding to
the four gauge groups $U(1)$, $SU(2)$, $SU(3)$
and $SU(5)$ respectively.
{}From \cite{PDG} we find
\begin{eqnarray}
\label{alpha1SM}
\alpha_{1}^{-1}(M_{Z}) & = & 58.85 \pm 0.10 \\
\alpha_{2}^{-1}(M_{Z}) & = & 29.794 \pm 0.048 \\
\alpha_{3}^{-1}(M_{Z}) & = & 8.55 \pm 0.37
\end{eqnarray}
where we have used the standard GUT
normalisation for $\alpha_{1}$.
For definiteness, we shall assume that
$\alpha_5^{-1}(M_Z) = 2$, although its
precise value is unimportant.
The first order RGEs \cite{GQW,RGEs,SMRGE}
are easily integrated and
give the results \cite{longpaper} shown in Fig.~\ref{alph4g}.
We see that there are no problems with Landau poles
below the Planck scale and our model is perturbatively consistent.

We can now use the RGEs to estimate upper
limits on the values of the Yukawa
couplings at the electroweak scale. This will lead to upper
limits on the masses of the fermions.
We do this by choosing initial values for the
Yukawa couplings at the Planck
scale and use the RGEs to see how they evolve,
as they are run down to the
electro-weak scale. Details of the RGEs we
used in the approximation of no
quark mixing are given in \cite{longpaper}.

\begin{figure}
\epsfxsize=\textwidth
\epsffile[100 100 440 500]{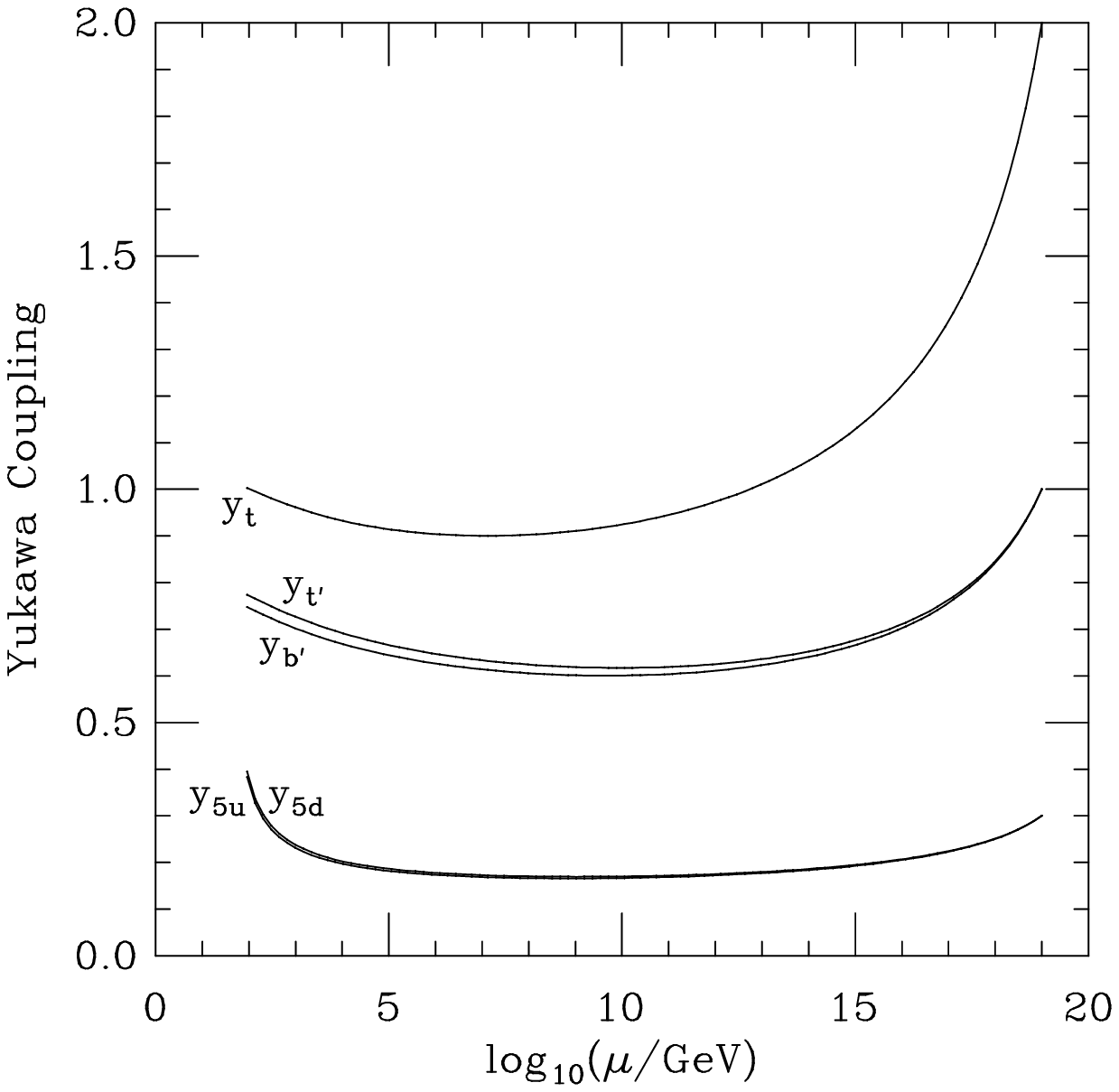}
\caption{}
%{An example of running Yukawa couplings
%for all fermions with a mass the same order of
%magnitude as the electroweak scale. The values were
%chosen at the Planck scale and run down to
%$M_Z$ so that all the fermions would
%have a mass allowed by current experimental limits.}
\label {yuk4g}
\end{figure}

We have chosen the low energy scale to be $M_Z$.
We observe fixed points
similar to the case for the top quark in the
SM \cite{hill}. However, the
Yukawa coupling for any fermion at $M_Z$
depends on the Yukawa couplings of
the other fermions. But there is an approximate
infrared fixed point limit on
$Y_2(S)=3y_t^2+3y_{t'}^2+3y_{b'}^2+5y_{5u}^2+5y_{5d}^2$
where we have used $5u$
and $5d$ to label the $SU(5)$ ``quarks''
as a generalisation of the naming of
the quarks. So one Yukawa coupling can be
increased at the expense of the
others. This limit is quite precise if there is only one strong
interaction at low energies such as QCD in the
SM \footnote{Detailed results
for a general number of heavy SM generations
are derived in \cite{SumM2}.}.
We observe numerically that $Y_2(S) \approx 7.5 \pm 0.3$
provided the Yukawa
couplings of the $t$, $t'$ and $b'$ quarks are greater than 1,
and that the Yukawa couplings of the $SU(5)$ ``quarks''
are less than the Yukawa couplings of the
$t$, $t'$ and $b'$ quarks, at the Planck scale.
The Yukawa couplings at the Planck scale in
fig.~\ref{yuk4g} have been chosen so that
$M_{t} \approx 170 \; \rm{GeV}$ and the
fourth generation quark masses are above
the current experimental limit of 130 GeV.
Also $M_{b'} \sim M_{t'}$ and
$M_{5u} \sim M_{5d}$ have been chosen so that there is only a small
contribution to the $\rho$ parameter.
Table~\ref{MaxMasses} gives the value of the Yukawa couplings
at $M_Z$ and the corresponding pole masses
including the technicolour
contribution to the VEV, as in eq.~(\ref{runmass}).
The relation between the
pole mass, $M_f$, and the running mass, $m_f$,
of a fermion $f$ is given by:
\begin{equation}
M_f=m_f(M_f) \left( 1+\frac{4}{3}\alpha_3(M_f) \right)
\end{equation}
for the quarks, and for the $SU(5)$ ``quarks'' by:
\begin{equation}
M_f=m_f(M_f) \left( 1+\frac{12}{5}\alpha_5(M_f) \right)
\end{equation}
The masses in table~\ref{MaxMasses} should
be considered upper limits on the
pole masses (the pole mass is identified with
the experimentally measured quantity for the $t$, $t'$ and $b'$
quarks) of the fermions for this particular choice
of Yukawa couplings at
the Planck scale. For other choices of Yukawa
couplings at the Planck scale we
could, for example,
increase the mass of the fourth generation
of quarks but this would have to be
compensated for by a reduction in
the mass of some of the other fermions.
These values for the masses are consistent with current experimental limits but
are not so high that all the new fermions could remain undetected for long.
In fact the fourth generation quark masses
may even be within the limits
of current accelerators.
However it is unlikely that the fermions
coupling to $SU(5)$ could be observed;
they would be confined inside $SU(5)$ ``hadrons'',
with a confinement scale
of order 400 GeV, and would have a
small production cross section at hadron
colliders. For this reason we consider the clearest
evidence for this model would come from the
detection of a fourth generation
quark. The masses of some fermions could be
increased, but not by much,
since this would mean a reduction in the mass
of other fermions. This means
that this model is consistent and relatively easy to test.

We also note that in \cite{longpaper}
we show that these fermion masses lead
to a fixed point value of the Higgs self-coupling,
which implies that the Higgs
particle has a mass of approximately 170 GeV.
This is the value we
assumed for the Higgs mass when discussing
the experimental limits on the $S$ parameter.

\section{Conclusion}
\label{Conclusion}

In answer to our title we would say that
it is possible to have another
generation of quarks without more leptons.
However, we have shown that even
with the smallest model of the types
we proposed (taking the additional gauge
group to be $SU(5)$), there are very severe
bounds from both precision
electroweak data and experimental lower limits
on the fourth generation quark
masses. We are clearly on the limit of agreement
with precision electroweak
data and this is sufficient to rule out the
models with an additional $SU(N)$
where $N>5$ since there would then be even
more $SU(2)$ doublets and so the
$S$ parameter would deviate from experiment by more than
2-3 standard deviations. In the case of
$N=5$ which we examined in detail we showed that
it is possible for the fourth
generation quarks to have masses above
the current experimental limits.
However, the upper limits from the infrared
quasi-fixed point limits on the
Yukawa couplings are not much above the
experimental lower bounds. When we
consider this together with the unnatural
requirement (in the sense that it
doesn't apply to the observed quarks)
that the quark masses be approximately
degenerate, it is clear that this model
can only be consistent with a very
limited choice of parameters. This is not
enough to conclude that the model
cannot be correct but more accurate
measurements of the $S$ parameter and
slightly higher limits on the masses
of a fourth generation of quarks would
certainly provide a definite conclusion.

\vspace{1cm}

\centerline{\bf Acknowledgements}
\vspace{.3cm}

We thank John Gunion, Hans Jensen, Victor Novikov,
David Sutherland and Misha
Vysotsky for helpful discussions.
This work has been supported in part by
INTAS Grant No.\ 93-3316 and PPARC Grant No.\ GR/J21231,
the British Council,
Cernfoelgeforskning and EF contract SC1 0340 (TSTS).

\newpage

\section*{Tables}
\begin{enumerate}
\item Left-handed fermions comprising an $SU(5)$
``quark'' generation. The electric charges are
in units of $\frac{1}{5}$ due to the charge quantisation rule.

\vspace{1cm}
\item Infrared fixed point Yukawa couplings and
corresponding pole masses (for $F_{\pi}=75$ GeV) for a
particular choice of Yukawa couplings at the Planck scale.
\end{enumerate}
\vspace{3cm}

\section*{Figures}
\begin{enumerate}
\item $\alpha^{-1}$ from $M_Z$ to the Planck scale for each group.
There are obviously no Landau poles so this model is self-consistent.

\vspace{1cm}
\item An example of running Yukawa couplings
for all fermions with a mass the same order of
magnitude as the electroweak scale. The values were
chosen at the Planck scale and run down to
$M_Z$ so that all the fermions would
have a mass allowed by current experimental limits.
\end{enumerate}

\vspace{6cm}

%\pagebreak
\newpage
\begin{table}
\caption{}
%{Left-handed fermions comprising an $SU(5)$
%``quark'' generation. The electric charges are
%in units of $\frac{1}{5}$ due to the charge quantisation rule.}
\begin{displaymath}
\begin{array}{|c|c|c|}
\hline
\rm{Representation\;under} & U(1)\;\rm{Representation} &
					\rm{Electric\;Charge} \\
SU(2) \otimes SU(3) \otimes SU(5) & \frac{y}{2} & Q \\ \hline
2,1,5 & -\frac{1}{10} & \left ( \begin{array}{c} \frac{2}{5} \\
-\frac{3}{5} \end{array} \right ) \\ \hline
1,1,\overline{5} & -\frac{4}{10} & -\frac{2}{5} \\ \hline
1,1,\overline{5} & \frac{6}{10} & \frac{3}{5} \\ \hline
\end{array}
\end{displaymath}
\vspace{6cm}
\label{SU5Quarks}
\end{table}

%\vspace{12cm}
%\newpage

%
\begin{table}
\caption{}
%{Infrared fixed point Yukawa couplings and
%corresponding pole masses (for $F_{\pi}=75$ GeV) for a
%particular choice of Yukawa couplings at the Planck scale.}
\begin{displaymath}
\begin{array}{|c|c|c|} \hline
$Fermion$ & $Yukawa Coupling at $M_Z$$ & $Pole Mass (GeV)$ \\ \hline
y_t     & 1.00 & 175    \\ \hline
y_{t'}  & 0.77 & 135    \\ \hline
y_{b'}  & 0.75 & 131    \\ \hline
y_{5u}  & 0.38 & 94     \\ \hline
y_{5d}  & 0.40 & 97     \\ \hline
\end{array}
\end{displaymath}
\label{MaxMasses}
\end{table}
%
%****************
\newpage
%****************
%*****************
%*****************
\end{document}